\documentclass[12pt]{article}
\usepackage{graphicx, amsmath,float}
\usepackage{multirow}
\usepackage{booktabs}
\usepackage{array}
\usepackage{lineno}
\usepackage{units}
\usepackage{color}
\usepackage{cite}
\usepackage{amsfonts}
\usepackage{mathtools}

\newlength{\dinwidth}
\newlength{\dinmargin}
\def\lapproxeq{\lower .7ex\hbox{$\;\stackrel{\textstyle                                                    
<}{\sim}\;$}}                                                    
\def\gapproxeq{\lower .7ex\hbox{$\;\stackrel{\textstyle                                                    
>}{\sim}\;$}}                                                                                                  
\def\bea{\begin{eqnarray}}                                                    
\def\eea{\end{eqnarray}}

\def\sh{\hat s}
\def\sh2{{\hat s}^2}

\newcommand{\be}{\begin{equation}}
\newcommand{\ee}{\end{equation}}

\begin{document}
                                                    
\titlepage                                                    
\begin{flushright}                                             
IRMP-CP3-25-43

\today \\                                                    
\end{flushright} 
\vspace*{0.5cm}
\begin{center}

{\Large \bf Phenomenological studies of exclusive heavy-quarkonium electroproduction  at~NLO}\\

\vspace*{0.5cm}
Chris A. Flett$^{a,b}$

\vspace*{0.5cm}  
\fontsize{10.47}{1}
$^a${\it Université Paris-Saclay,
CNRS, IJCLab, 91405 Orsay, France}\\
$^b${\it Centre for Cosmology, Particle Physics and Phenomenology (CP3), Universit\'e Catholique de Louvain, Chemin du Cyclotron, Louvain-la-Neuve,
B-1348, Belgium}                                         
                                      
\vspace*{1cm}                                                    

\begin{abstract} 
\vspace*{0.2cm} 
Using the next-to-leading order~(NLO) coefficient functions for exclusive electroproduction of heavy vector mesons derived in our previous work, we perform various phenomenological studies of exclusive electroproduction in $ep$ collisions relevant for both the existing measurements from HERA, and the forthcoming Electron-Ion Collider~(EIC). We compare our cross-section results to HERA data across a broad range of photon virtualities $Q^2$ and $\gamma^* p$ centre-of-mass energies, provide predictions for upcoming EIC measurements and conclude with a discussion on the necessity of resumming logarithmically enhanced contributions in $J/\psi$ electroproduction.
\end{abstract}     
\vspace*{0.5cm}                                                                                                
\end{center} 

\section{Introduction}
The exclusive production of heavy vector mesons has long been a compelling subject of study, serving as a promising approach to understanding the small-$x$ behavior of the gluon Parton Distribution Function~(PDF) over a wide range of resolution scales. Initially measured in fixed-target experiments and then
in electron-proton ($ep$) Deep-Inelastic Scattering~(DIS) at HERA, before more recently in hadron-hadron ultraperipheral collisions at the LHC, these observations offer valuable insights into the quarkonium-production mechanism and serve as precise probes at the forefront of small-$x$ saturation physics. 

The exclusive $J/\psi$ electroproduction, $\gamma^* p \rightarrow J/\psi p$, has been measured through di-lepton decays at HERA in $ep$ collisions within a wide range of photon virtualities $Q^2$, up to $100~ \text{GeV}^2$~\cite{H1:1996gwv, ZEUS:2004yeh, H1:2005dtp}. The corresponding photoproduction process, with $\langle Q^2 \rangle \approx 0$, has also been measured~\cite{H1:1996kyo, H1:2000kis, ZEUS:2004yeh, H1:2005dtp, H1:2013okq}, or extracted from ultraperipheral collision data at the LHC~\cite{LHCb:2014acg, ALICE:2014eof, ALICE:2018oyo, LHCb:2018rcm, ALICE:2023mfc}. There are, however, no data so far from experiment for exclusive $\Upsilon$ electroproduction,  $\gamma^* p \rightarrow \Upsilon p$, beyond the photoproduction limit~\cite{ZEUS:1998cdr, H1:2000kis, ZEUS:2009asc,LHCb:2015wlx, CMS:2018bbk}. Moving forward, the upcoming Electron-Ion Collider~(EIC) will provide a lever arm in the virtuality $Q^2$, enabling measurements of $\Upsilon$ electroproduction with off-shell kinematics for the first time. This will include higher virtualities, although the coverage in $Q^2 + M_{V}^2$ and anticipated event-count rates will be lower than for $J/\psi$ due to the heavier mass $M_V$ of $\Upsilon$~\cite{AbdulKhalek:2021gbh}. 

In a previous work~\cite{Flett:2021ghh}, we derived the coefficient functions for exclusive heavy vector-meson electroproduction at next-to-leading order~(NLO) within the framework of collinear factorisation~(CF), with the transition of the open heavy quark-antiquark pair to a bound heavy vector meson made within Non-Relativistic QCD~(NRQCD)~\cite{Hoodbhoy:1996zg}. In this work, we build upon these results to perform detailed phenomenological analyses, confronting NLO CF + NRQCD predictions with the experimental data. Sec.~\ref{overview} gives an overview of our computation and explains our set up within CF. Using these results, Sec.~\ref{results} then compares our \textcolor{black}{predictions} \textcolor{black}{for various experimentally measured observables} in  $\gamma^* p \rightarrow J/\psi p$ to data from HERA, and discusses the experimental feasibility of similar measurements at the EIC for both $J/\psi$ and $\Upsilon$ production. We also identify the need for resummation of a particular class of logarithmically-enhanced terms and assess their present phenomenological relevance through comparison with data at the highest virtualities for $J/\psi$ electroproduction. Finally, Sec.~\ref{conc} gathers our conclusions.

\section{Overview of framework}
\label{overview}
In~\cite{Ivanov:2004vd}, the NLO coefficient functions for the exclusive photoproduction of heavy quarkonium $V$, $\gamma p \rightarrow V p$, were computed using dispersion relations within CF. Explicitly, the imaginary parts of the diagrams which have a discontinuity in the $s$-channel were calculated, together with the real part of the amplitude and the $u$-channel contribution using a dispersion relation\footnote{For the quark subprocess, which contributes at the 1-loop level, the dispersion integral was found to be readily convergent, allowing the real part to be directly restored. In contrast, for the gluon contribution at 1-loop, it was necessary to construct a once-subtracted dispersion relation.}. In our approach for electroproduction~\cite{Flett:2021ghh}, we work too within CF but directly compute the real and imaginary parts of the coefficient functions using integral reduction.\footnote{We have checked that these agree numerically with the coefficient functions computed via a similar approach in~\cite{Chen:2019uit}.} In this section, we give a brief overview of our computation and framework.

We generate all LO and NLO Feynman diagrams using \texttt{QGRAF}~\cite{Nogueira:1991ex}; the diagrams compatible with the external colour and kinematic constraints are processed in \texttt{FORM}~\cite{Ruijl:2017dtg} in an arbitrary linear covariant gauge. The appropriate quark or gluon Generalised Parton Distribution~(GPD) projector is applied to each diagram, together with the quarkonium-spin projection, and the resulting Dirac traces are computed in $d =  4 - 2\epsilon$ space-time dimensions. Due to the external colour and kinematic constraints, the integral structures obtained contain in general linearly-dependent propagators. This linear dependence is first removed following a generalised partial-fractioning procedure facilitated by the package \texttt{Apart}~\cite{Feng:2012iq}, before Integration-by-Parts~(IBPs) identities are employed to reduce the integrals to a set of so-called master integrals, achieved here with \texttt{Reduze 2}~\cite{vonManteuffel:2012np}.

After substitution of these master integrals, the bare amplitude contains simple poles in $\epsilon$. We renormalise the gluon, heavy-quark wavefunction and the heavy-quark mass in the on-shell scheme, while the strong-coupling constant is renormalised with light flavours treated in the $\overline{{\rm MS}}$ scheme and with the heavy-quark loop of the gluon self energy subtracted at zero momentum. After incorporating such counter-terms, the  amplitude is free from ultraviolet~(UV) divergences but poles in $\epsilon$ still remain in both the bare quark and gluon one-loop coefficient functions, which however can be identified as mass singularities and can thus be suitably absorbed into the definition of the bare GPDs through a consistent mass-factorisation renormalisation procedure. The addition of the mass factorisation and UV counter-terms renders renormalised, finite NLO coefficient functions that may be convoluted with renormalised, finite GPDs to obtain the exclusive electroproduction amplitude to NLO accuracy.

\begin{figure}[htbp]
\begin{center}
\includegraphics[width=0.29\textwidth]{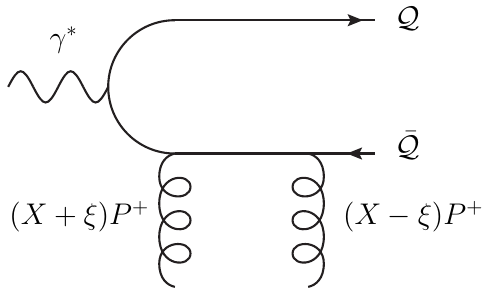}
\quad
\includegraphics[width=0.26\textwidth]{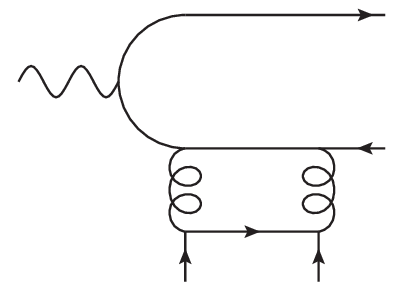}
\quad
\includegraphics[width=0.26\textwidth]{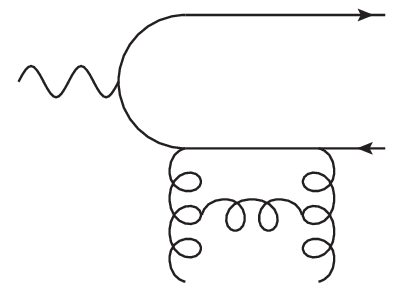}
\caption{\sf{From left to right: Example Feynman diagrams for the LO gluon, NLO quark and NLO gluon-initiated subprocesses. Here, the average of the incoming and outgoing proton momenta, $P^\mu$, defines the collinear direction. The momentum fractions carried by the incoming and outgoing partons along the light-cone component $P^+$ are $X+\xi$ and $X-\xi$, respectively. All diagrams were produced using JaxoDraw~\cite{Binosi:2003yf}.}}
\label{fig:diags}
\end{center}
\end{figure}

The amplitude for the exclusive electroproduction of heavy quarkonium off unpolarised protons, $\gamma^* p \rightarrow V p$,
reads
\begin{align}
\begin{split}
\mathcal{T}_{ss'}^{\mu \nu} =& 
 -g^{\mu \nu}_\perp \left(\frac{\langle O_1 \rangle_V}{m_{\mathcal Q}^3} \right)^{1/2} \int_{-1}^1 \frac{ \text{d} X}{X} \left[ \sum_q F_{ss'}^q(X,\xi) C_{\perp,q}\left(\frac{\xi}{X}, \frac{m_{\mathcal Q}^2}{Q^2} \right) + \frac{F_{ss'}^g(X,\xi)}{X}C_{\perp,g}   \left(\frac{\xi}{X}, \frac{m_{\mathcal Q}^2}{Q^2} \right) \right]\\
&+\ell^{\mu \nu} \left(\frac{\langle O_1 \rangle_V}{m_{\mathcal Q}^3} \right)^{1/2} \int_{-1}^1 \frac{\text{d} X}{X} \left[ \sum_q F_{ss'}^q(X,\xi) C_{L,q}\left(\frac{\xi}{X}, \frac{m_{\mathcal Q}^2}{Q^2} \right) + \frac{F_{ss'}^g(X,\xi)}{X}C_{L,g}  \left(\frac{\xi}{X}, \frac{m_{\mathcal Q}^2}{Q^2} \right)  \right]. 
\end{split} \label{eq:HVMPartonFac}
\end{align}
Here, $X$ is the average momentum fraction of the parton in the proton $p$, and the presence of the skewness variable $\xi$ underlies the factorisation in terms of quark and gluon GPDs, denoted $F^q_{ss'}$ and $F^g_{ss'}$ respectively, with $s,s'$ labelling proton spins. 
$C_{\perp,k}$ ($C_{L,k}$) is the quark $(k=q)$ or gluon $(k=g)$ transverse (longitudinal) coefficient function to NLO, see~\cite{Flett:2021ghh}, expressed in terms of dimensionless ratios $\xi/X$ and $m_{\mathcal Q}^2/Q^2$, where $m_{\mathcal Q}$ is the heavy-quark mass. Exemplary Feynman diagrams contributing to these coefficient functions to NLO are shown in Fig.~\ref{fig:diags}. 
The Lorentz indices $\mu$ and $\nu$ are
those of the incoming photon and outgoing quarkonium respectively; the tensors $g_{\perp}^{\mu \nu}$ and $\ell^{\mu \nu}$ denote the perpendicular metric and longitudinal projectors, see~\cite{Flett:2021ghh} for details. Additional dependences of the GPDs on the factorisation scale $\mu_F$ and on the \textcolor{black}{four-momentum transfer $t$} have been suppressed for clarity, as has the $\mu_F$ and renormalisation-scale $\mu_R$ dependence of the coefficient functions. \textcolor{black}{In phenomenological studies, the dependence of GPDs on $X$ and $\xi$ is often assumed to factorise from the $t$-dependence.}\footnote{The minimum value of the momentum transfer is $t = t_{\rm min} = 4\xi^2 m_p^2/(1-\xi^2)$, where $m_p$ is the mass of the proton. \textcolor{black}{In the limit $W^2 \gg Q^2+M_V^2$, the skewness $\xi \approx (Q^2+M_V^2)/(2W^2)$, with $W$ denoting the $\gamma^* p$ centre-of-mass energy}. For such energies reachable at the EIC and HERA, the value of $t_{\rm min}$ is therefore practically zero.} The long-distance matrix element, $\langle O_1 \rangle_V$, represents the colour-singlet non-perturbative matrix element for the $\mathcal Q \bar{\mathcal Q} \rightarrow V$ transition. It can be determined via potential models or the di-lepton decay width of the heavy quarkonium, as described in~\cite{Ivanov:2004vd}.

As measurements of heavy-quarkonium production from unpolarised targets probes only the charge-conjugation-even quark GPD, in the above we replace $\Sigma_q F^q_{ss'}$ with the quark-singlet GPD $F^S_{ss'}/2$. \textcolor{black}{We construct the quark-singlet and gluon GPD using the so-called Shuvaev transform~\cite{Shuvaev:1999ce,Shuvaev:1999fm}, which relates GPDs at small skewness to the corresponding forward quark $q(x)$ and gluon $g(x)$ PDFs. For unpolarised protons we consider the parton-helicity conserving matrix elements
$F_{ss'}^j$ with $j=q,g$. In this work, we do not consider the off-diagonal proton-helicity flip contributions with $s \neq s'$ and so the spin indices can be suppressed and the resulting scalar GPDs can simply be denoted by $F^j$. These GPDs are then given by\footnote{\textcolor{black}{The prefactor $\sqrt{1-\xi^2}$ is a kinematic factor originating from the spinor normalisation of the proton states for equal helicities, $s=s’$, when reducing the operator definition of the GPDs to scalar functions.}}}
\begin{equation}
\begin{split}
F^q(X,\xi) &= \sqrt{1-\xi^2}\int_{-1}^{1} dx'\,
\mathcal{K}_q(X,\xi;x')
\frac{d}{dx'}\left(\frac{q(x')}{|x'|}\right),\\
F^g(X,\xi) &= \sqrt{1-\xi^2}\int_{-1}^{1} dx'\,
\mathcal{K}_g(X,\xi;x')\,
\frac{d}{dx'}\left(\frac{g(x')}{|x'|}\right),
\label{Shuvaevt}
\end{split}
\end{equation}
where the Shuvaev kernels are defined by
\begin{equation}
\begin{split}
\mathcal{K}_q(X,\xi;x') &=
\frac{2}{\pi}\,\mathrm{Im}\,\int_{0}^{1}
\frac{d\alpha}{y(\alpha)\sqrt{1-y(\alpha)x'}},\\
\mathcal{K}_g(X,\xi;x') &=
\frac{2}{\pi}\,\mathrm{Im}\int_{0}^{1}
\frac{d\alpha \bigl[X+\xi(1-2\alpha)\bigr]}{y(\alpha)\sqrt{1-y(\alpha)x'}},
\end{split}
\end{equation}
and with the auxiliary function 
\begin{equation}
y(\alpha)=\frac{4\alpha(1-\alpha)}{X+\xi(1-2\alpha)}.
\label{ys}
\end{equation}
This transform leverages the fact that as $\xi \rightarrow 0$ (and at $t = 0$), the conformal moments of the GPD become equal to the known Mellin moments of the PDF. Due to the polynomiality, even for $\xi \neq 0$ the conformal moments can be derived from the Mellin moments with $O(\xi)$ accuracy. It is therefore possible to reconstruct the complete GPD at small $\xi$ from its known moments. Importantly, the analytic continuation of such moments to the complex $N$ plane may induce singularities and thereby impact the reliability of the transform. It was argued in~\cite{Martin:2008gqx} that such singularities cannot however occur in the right half of the $N>1$ plane in which lies the DGLAP region, $|X| > \xi$. For this reason we employ the Shuvaev transform in this region only, evaluating eqn.~\eqref{eq:HVMPartonFac} using the imaginary part of the coefficent function and restoring the real part at amplitude level using the well-known derivative analyticity relation~\cite{Gribov:1968uy, Ryskin:1995hz}.

We note that, in the gauge-invariant sum of all Feynman diagrams, the apparent end-point singularities at $X=\pm \xi$
associated with the gluon GPD projector must cancel. In our approach, this cancellation is verified by expanding the amplitude around $X=\pm \xi$ and demonstrating numerically that the singular terms vanish once all contributions are summed. These apparent singularities occur at kinematic points where an active gluon carries vanishing light-cone momentum, see Fig.~\ref{fig:diags}, and are not related to any physical intermediate state becoming on-shell. Rather, they arise only in spurious denominators from gauge-dependent terms in the gluon propagator in the axial gauge.  

Equation~\eqref{eq:HVMPartonFac} gives the exclusive electroproduction amplitude manifestly decoupled into the transverse and longitudinal degrees of freedom through the terms proportional to $g_{\perp}^{\mu \nu}$ and $\ell^{\mu \nu}$. Contractions of eqn.~\eqref{eq:HVMPartonFac} with the explicit transverse ($\pm$) and longitudinal ($0$) polarisation vectors of the photon and quarkonium give the helicity amplitudes. The helicity-flip contributions are not relevant here and we thus evaluate the (non-vanishing) helicity amplitudes
\begin{equation}
\mathcal{T}_{ss'}^{\pm \pm} = \varepsilon_{\pm\mu}^{\gamma^*} \mathcal{T}_{ss'}^{\mu \nu} \varepsilon_{\pm\nu}^{*V}\,\,\,\,\,{\rm and}\,\,\,\,\, \mathcal{T}_{ss'}^{0 0} = \varepsilon_{0\mu}^{\gamma^*} \mathcal{T}_{ss'}^{\mu \nu} \varepsilon_{0\nu}^{*V},
\label{pol}
\end{equation}
corresponding to the photon and quarkonium helicity states $\left\{ij\right\}=\left\{++,--,00\right\}$.

The $\gamma^* p \rightarrow V p$ two-body $t$-differential cross section at $t=0$ is readily obtained from the amplitudes, eqn.~\eqref{pol}, as usual from 
\begin{equation}
\frac{d \sigma}{dt}(\gamma^* p \rightarrow V p)|_{t=0} = \frac{1}{16 \pi \lambda(W^2, Q^2, m_p^2)} ~\overline{\sum}_{i,j={\pm,0}}\,\overline{\sum}_{s,s'} |\mathcal T_{ss'}^{ij}|^2,
\end{equation}
where the sums include the initial-state averaging of the incoming proton spins $(s,s')$ and the photon transverse and longitudinal helicities $(\pm,0)$. 
Here, $\lambda(x,y,z) = x^2 + y^2 + z^2 - 2xy - 2xz - 2yz$ is the well-known K\"{a}llen function.
The total $t$-integrated cross section \textcolor{black}{is obtained by assuming an exponential $t$-dependence of the GPD, which leads to the commonly adopted form $d \sigma/dt \propto e^{-B|t|}$ at the cross-section level, namely}
\begin{equation}
    \sigma(\gamma^* p \rightarrow V p) = \int_{-\infty}^{t_{{\rm min}}} dt \frac{d \sigma}{dt}\bigg|_{t=0}~e^{B(W)t} = \frac{1}{B(W)} \frac{d \sigma}{dt}\bigg|_{t=0},
\end{equation}
where the $B$-slope $B(W) = B_0 + 4\alpha_{{\mathbb{P}}}' \ln(W/90) \,\,{\rm GeV}^{-2}$. 
The intercept $B_0 = 4.9~(4.63) \,{\rm GeV}^{-2}$ for $J/\psi$ ($\Upsilon$) production and $\alpha_{\mathbb{P}}' = 0.06\, {\rm GeV}^{-2}$. This parametrisation grows more slowly with $W$ than that used by H1~\cite{H1:2013okq}; the value of $B_0$ is compatible with fits to the $t$ dependence of elastic $J/\psi$ photoproduction data on the proton at HERA~\cite{H1:2000kis,H1:2013okq}, while the Pomeron trajectory slope $\alpha_\mathbb{P}'$ is fixed by Model 4 of~\cite{Khoze:2013dha}, \textcolor{black}{which fits a wider variety of diffractive data taken in a small momentum-transfer window $|t| < 1$ GeV$^2$}. Moreover, as shown in~\cite{Kowalski:2006hc}, the $B$ vs. $Q^2+M_V^2$ distribution is rather flat for heavy vector-meson production, considered here,
and so we use the same $B$-slope parametrisation in all of our numerical results presented in Sec.~\ref{results}. 

\section{Results}
\label{results}
In this section, we present and discuss our results for the $V=J/\psi$ and $\Upsilon$ electroproduction process on the proton, $\gamma^* p \rightarrow V p,$ as a function of the photon virtuality $Q^2$ and centre-of-mass energy $W$ of the $\gamma^* p$ subprocess.  The kinematical capabilities of the EIC for exclusive heavy vector-meson production are shown in~\cite{AbdulKhalek:2021gbh}. The simulated distribution of events for $J/\psi$ and $\Upsilon$ production in the two different beam-energy configurations of the EIC, assuming an integrated luminosity of 100~fb$^{-1}$, are given in the $(x, Q^2)$ plane.\footnote{Here, the skewness parameter \textcolor{black}{$\xi \approx (Q^2+M_V^2)/(2W^2)$}  $=x/2$ for \textcolor{black}{$W^2 \gg Q^2+M_V^2$}.}  

\begin{figure}[htbp]
\begin{center}
\includegraphics[width=0.484\textwidth]{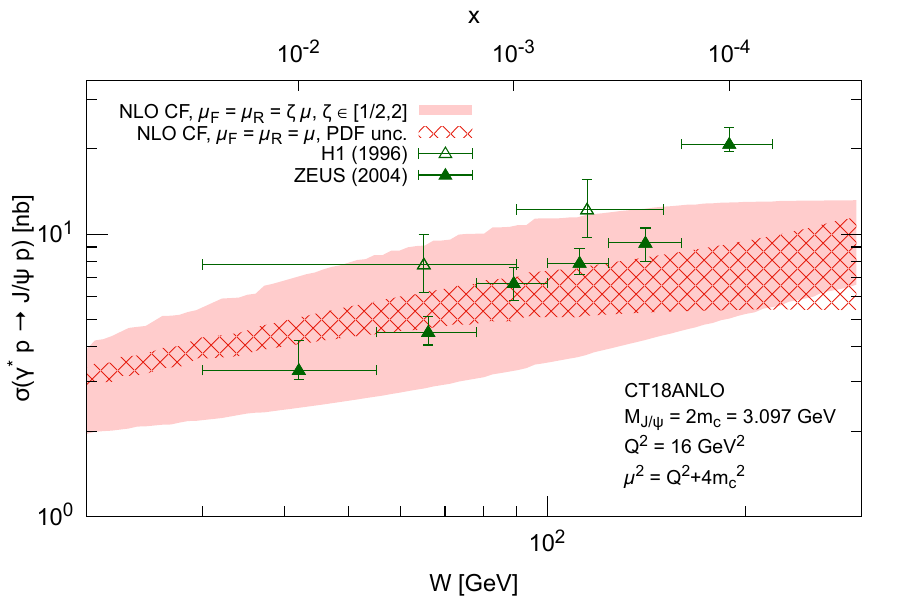}
\quad
\includegraphics[width=0.484\textwidth]{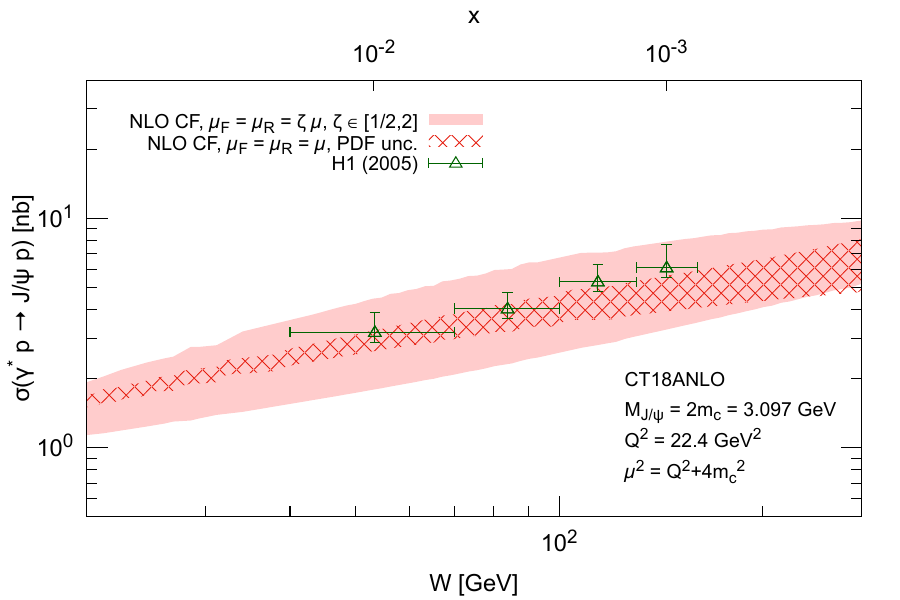}
\caption{\sf{Postdictions for exclusive $J/\psi$ electroproduction as a function of $W$ for $\langle Q^2 \rangle = 16$~GeV$^2$ (left) and $22.4$~GeV$^2$ (right), and compared to experimental measurements from HERA~\cite{H1:1996gwv, ZEUS:2004yeh, H1:2005dtp}. The shaded band represents the $\mu_{F,R}$ scale-variation uncertainty about the central scale $\mu^2 = Q^2 + 4m_c^2$, while the hatched band represents the NLO PDF uncertainty.}}
\label{fig:chris_2}
\end{center}
\end{figure}

In Fig.~\ref{fig:chris_2} we show the total $t$-integrated cross section of exclusive $J/\psi$ electroproduction as a function of the $\gamma^* p$ centre-of-mass energy $W$ at $\langle Q^2 \rangle$ = 16 and 22.4~GeV$^2$, using Shuvaev-transformed CT18ANLO PDFs and compared to existing experimental data from HERA~\cite{H1:1996gwv, ZEUS:2004yeh,H1:2005dtp}. The upper axis labels the typical momentum fraction $x \approx 2 \xi$ that is probed as a function of $W$.

The shaded band corresponds to a scale variation with $\mu_{R}=\mu_{F}\in\{\mu/2,\mu,2\mu\}$ around the central scale $\mu=\sqrt{Q^2+4m_c^2}$. The choice $\mu_R = \mu_F$ is chosen for two reasons. First, it corresponds to the BLM scale prescription\cite{Brodsky:1982gc} and eliminates terms proportional to $\beta_0$ in the hard-scattering gluon coefficient function at NLO. Secondly, the one-loop quark-loop insertion in the gluon propagator contributes to both the running of the strong coupling, through $\beta_0$, and the virtual correction in the DGLAP splitting kernel, through $\delta(1-z) \propto \beta_0/2$. These two contributions describe the same physical effect at different momentum scales. To avoid a mismatch or a double counting, one can identify $\mu_R = \mu_F$. The hatched band represents the PDF uncertainty at the central scale $\mu$, propagated using the standard Hessian eigenvector method.
Note that, for \textcolor{black}{$W^2 \gg Q^2+M_V^2$}, the NLO amplitude develops sensitivity to a double-logarithmic enhancement $\propto \ln (1/\xi) \ln(\bar{\mu}^2/\mu_F^2)$, where $\bar{\mu}^2 = (Q^2 + 4m_c^2)/4$.\footnote{This scale corresponds to that used for the lower boundary of the shaded bands in Fig.~\ref{fig:chris_2}.} In fact, such terms belong to a family of double logarithms $\propto \alpha_s^n \ln^n (1/\xi) \ln^n(\bar{\mu}^2/\mu_F^2)$. \textcolor{black}{These logarithms} are intrinsically resummed in the framework of $k_t$ factorisation, and this resummation can be effectively mimicked in CF through a judicious choice of factorisation scale, at least at NLO.\footnote{See~\cite{Flett:2024htj} for details.} However, as the kinematic reach of HERA and EIC is not in the regime of asymptotically large $W$, a resummation of these logarithms would not guarantee to capture all numerically leading corrections at the realistically accessible values of $W$. In the left panel, the central prediction agrees most favourably with the more up-to-date dataset, however the EIC is anticipated to provide more statistics and resolve the slight tension between (and discrepancies within) the datasets. In particular, the data point at $W = 189$~GeV deviates by around a factor of two from the other data lying in this $\langle Q^2 \rangle$ bin. Indeed, as shown in~\cite{AbdulKhalek:2021gbh}, the EIC is expected to deliver a statistically meaningful sample of events at $Q^2 = 16$~GeV$^2$ and $22.4$~GeV$^2$ for $x \gapproxeq 10^{-4}$.

\begin{figure}[htbp]
\begin{center}
\includegraphics[width=0.484\textwidth]{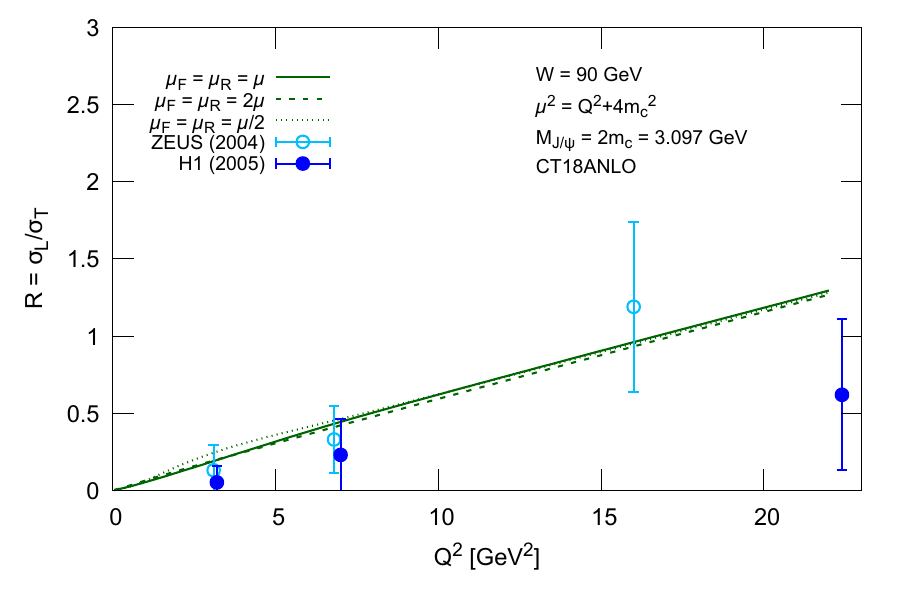}
\quad
\includegraphics[width=0.484\textwidth]{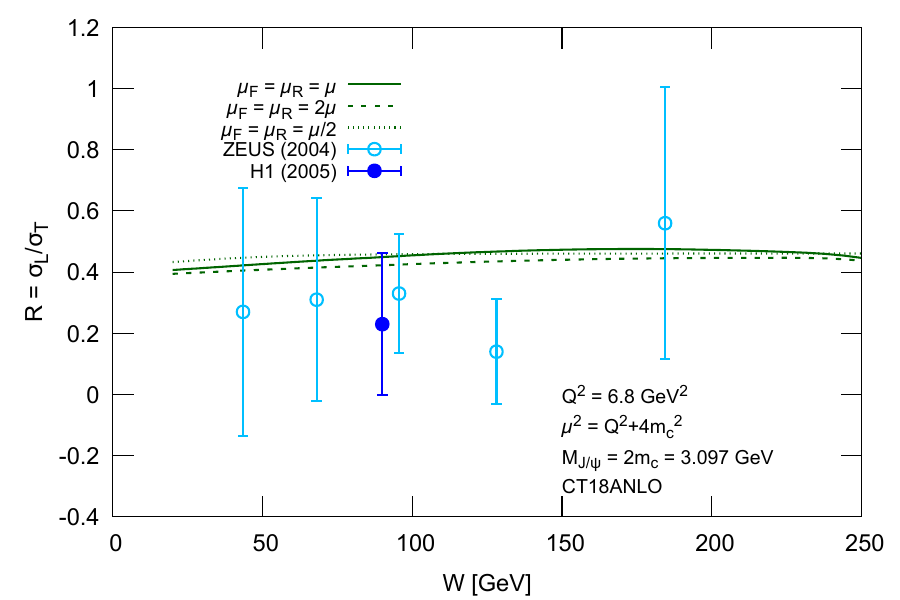}
\caption{\sf{Predictions of $R = \sigma_L/\sigma_T$, the ratio of the cross sections due to longitudinally-polarised photons and transversely-polarised photons, as a function of $Q^2$ (left) and $W$ (right) for exclusive $J/\psi$ electroproduction, and compared to experimental measurements from HERA~\cite{ZEUS:2004yeh, H1:2005dtp}.}}
\label{fig:chris_3}
\end{center}
\end{figure}
\textcolor{black}{In Fig.~\ref{fig:chris_3}, we show predictions for the ratio $R$ of the longitudinally-polarised cross section $\sigma_L$ to the transversely-polarised cross section $\sigma_T$ as a function of $Q^2$ and $W$, compared to data from both ZEUS~\cite{ZEUS:2004yeh} and H1~\cite{H1:2005dtp}. We show the same scale variation around the central value $\mu = \sqrt{Q^2+4m_c^2}$ as used in Fig.~\ref{fig:chris_2}. For increasing $Q^2$, the longitudinally-polarised photon couples more strongly to the small-size $c\bar{c}$ pair. This is shown in the left panel of Fig.~\ref{fig:chris_3} where the increasing relative contribution of $\sigma_L$ compared to $\sigma_T$ is quantified. The effective power growth in the $W$ dependence of $\sigma_L$ and $\sigma_T$ is similar and hence the predictions in the right panel of Fig.~\ref{fig:chris_3} show a flat behaviour.
As shown in both cases, \textcolor{black}{the ratio exhibits a milder dependence on the factorisation and renormalisation scales compared to the more pronounced scale dependence of the individual total cross sections shown in Fig.~\ref{fig:chris_2}}. These results motivate further data-taking at the EIC, which will provide greater statistics and help resolve, for example, the discrepancy between the datasets at large $Q^2$.}

$\Upsilon$ production, on the other hand, samples larger momentum fractions $x$ and higher virtualities $Q^2$ due to its larger mass. This results in lower event yield rates and access to a more restricted kinematic region, even at the maximum collider energy of the EIC. 
In its highest-energy configuration, the event count is three orders of magnitude lower than for $J/\psi$ in the photoproduction bin, which extends down in $x$ to a few units of $10^{-3}$. Outside of this bin, the event-count rate is very low~\cite{AbdulKhalek:2021gbh}. Meanwhile, the lowest-energy configuration essentially populates only the photoproduction bin in the valence $x$ region. Due to its larger mass, one can test factorisation at significantly larger scales, of the order $\sim O(M_\Upsilon^2)$. However, as noted, the $\Upsilon$ production yields drop sharply with increasing $Q^2$ and so the leverage of the EIC in this variable is rather limited for this observable. As a result, prospects for studying scale evolution at the EIC over a broad range of $Q^2$ are at best modest, especially when compared to the case of $J/\psi$. Any data will therefore likely be sparse and exhibit large uncertainties. \textcolor{black}{Nevertheless}, they will complement the measurements already published by HERA and the LHC in the photoproduction bin; see~\cite{Flett:2021fvo} for the corresponding integrated cross section for $\Upsilon$ photoproduction as a function of $W$ and the available data.

\begin{figure}[htbp]
\begin{center}
\includegraphics[width=0.6465\textwidth]{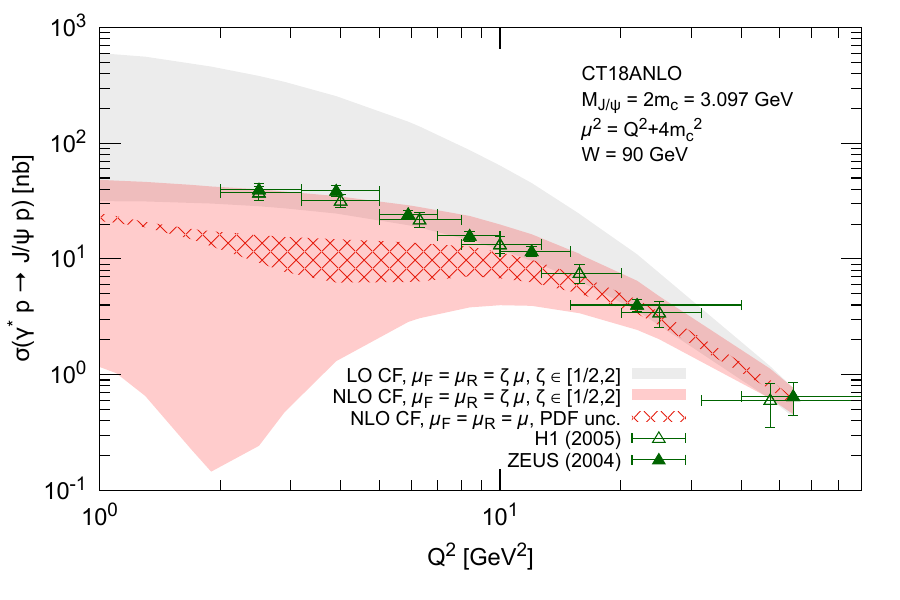}
\caption{\sf{The exclusive $J/\psi$ electroproduction cross section as a function of $Q^2$ for a fixed centre-of-mass energy $W$ = 90 GeV and compared to the data from ZEUS~\cite{ZEUS:2004yeh} and H1~\cite{H1:2005dtp}, using  results in~\cite{Flett:2021ghh} and Shuvaev-transformed input CT18ANLO PDFs~\cite{Hou:2019efy}. The grey shaded band represents the simultaneous $\mu_{F,R}$ scale variation uncertainty of the LO result, the light-red shaded band that of the NLO result, while the hatched band gives the propagation of the NLO PDF uncertainty to the cross-section level.
}}
\label{fig:chris_1}
\end{center}
\end{figure}

In Fig.~\ref{fig:chris_1}, we show instead the exclusive $J/\psi$ electroproduction cross section in bins of $Q^2$ at a fixed centre-of-mass energy, $W = 90~\text{GeV}$, of the $\gamma^* p$ pair. The grey (light-red) shaded band corresponds to the use of CF at LO (NLO), with Shuvaev-transformed CT18ANLO PDFs~\cite{Hou:2019efy}. The NLO PDF uncertainty propagated to the cross section is shown again via the hatched band. The NLO prediction is in agreement with the data from both the H1 and ZEUS collaborations \textcolor{black}{at $W$=90 GeV} over a wide range of $\langle Q^2 \rangle$.
We have checked that the choice of input NLO PDFs amongst CT18A~\cite{Hou:2019efy}, MSHT20~\cite{Bailey:2020ooq} and NNPDF3.1~\cite{NNPDF:2017mvq} has the largest effect at the lowest $Q^2$, where the choice of the initial condition of the DGLAP evolution is felt\textcolor{black}{; for larger $Q^2$} this effect washes out and the predictions based on each PDF set agree at or below the percent level. That is to say, the prediction in the highest $Q^2$ bin exhibits a small factorisation-scale dependence and is essentially independent of the choice of the input PDF.

At $Q^2 \gapproxeq 20\,\text{GeV}^2$, accessible at the EIC, the perturbative convergence of exclusive $J/\psi$ electroproduction manifests itself well with the LO and NLO predictions in line with each other. At smaller $Q^2$, however, the perturbative convergence deteriorates and we observe the well-known scale-stability problem of exclusive $J/\psi$ photoproduction at NLO in CF~\cite{Ivanov:2004vd}. The interplay between quark and gluon contributions to the total amplitude in standard CF is rather complicated in this small-$Q^2$ regime, $Q^2 \lapproxeq m_{c}^2$, see~\cite{Eskola:2022vpi, Flett:2024htj}. We find however that the gluon amplitude quickly becomes dominant for $Q^2 \gg m_{c}^2$ and, consequently, the forthcoming enhanced statistics and increased data coverage from the EIC will allow for refined and improved constraints on gluon parton distribution function extractions at intermediate resolution scales. This is reflected in both Figs.~\ref{fig:chris_2} and \ref{fig:chris_1} where, for $Q^2 \gg m_c^2$, the width of the hatched band is driven by the magnitude and uncertainty of the gluon PDF, whereas for $Q^2 \lapproxeq m_c^2$ this is no longer the case due to the delicate interplay of the contributing terms discussed above. The behaviour of the lower boundary of the light-red shaded band around $Q^2 \approx 2 ~\text{GeV}^2$ in Fig.~\ref{fig:chris_1} is due to the rapid decrease of the imaginary part of the NLO amplitude in this region, which leads to a sign change relative to the LO imaginary part for $Q^2 \lapproxeq 2 ~\text{GeV}^2$, in qualitative agreement with the observations of~\cite{Chen:2019uit}.

We end this section by remarking that, in the large $Q^2 \gg m_{\mathcal Q}^2$ limit, the transverse and longitudinal gluon (quark) coefficient functions contain logarithms in the ratio $Q^2/m_{\mathcal Q}^2$ up to order two (one):
\begin{align}
{\rm lim}_{Q^2 \gg m_{\mathcal Q}^2} C_{\perp,q}^{(1)}\left(\frac{\xi}{X}, \frac{m_{\mathcal Q}^2}{Q^2}\right) &\supset \ln(Q^2/m_{\mathcal Q}^2),\nonumber \\
{\rm lim}_{Q^2 \gg m_{\mathcal Q}^2} C_{\perp,g}^{(1)}\left(\frac{\xi}{X}, \frac{m_{\mathcal Q}^2}{Q^2}\right) &\supset \ln^2(Q^2/m_{\mathcal Q}^2),
\end{align}
and 
\begin{align}
{\rm lim}_{Q^2 \gg m_{\mathcal Q}^2} C_{L,q}^{(1)}\left(\frac{\xi}{X}, \frac{m_{\mathcal Q}^2}{Q^2}\right) &\supset \ln(Q^2/m_{\mathcal Q}^2),\nonumber \\
{\rm lim}_{Q^2 \gg m_{\mathcal Q}^2} C_{L,g}^{(1)}\left(\frac{\xi}{X}, \frac{m_{\mathcal Q}^2}{Q^2}\right) &\supset \ln^2(Q^2/m_{\mathcal Q}^2). 
\end{align}

The logarithmic enhancements proportional to $\ln(Q^2/m_{\mathcal Q}^2)$ are characteristic of calculations in the Fixed-Flavour-Number Scheme (FFNS), where heavy quarks are produced perturbatively in the hard subprocess rather than included as partons in the initial-state DGLAP evolution. We work within such a framework here, and stress in particular that in the asymptotic $Q^2 \gg m_{\mathcal Q}^2$ limit, the dominating gluon contribution exhibits quadratic growths in this logarithm.
Such double-logarithmic terms arise from the variant of the right-most Feynman diagram in Fig.~\ref{fig:diags} in which the upper gluon cell is replaced by a heavy-quark cell with an intermediate gluon propagator at the top of the cell.\footnote{Schematically, one logarithm arises from the usual DGLAP evolution in the transverse momentum of the $t$-channel quark propagators, while the second comes from the integration over the longitudinal momentum fraction, with limits involving the ratio $Q^2/m_{\mathcal Q}^2$.} In~\cite{Kirschner:1983di}, an analogous class of terms was resummed using the infrared evolution equation~(IREE). A similar approach could be adopted here, using the IREE in the region between the hard scale $Q^2$ and the infrared cutoff $m_{\mathcal Q}^2$, supplementing it with the usual DGLAP evolution to account for the single logarithms in $Q^2/m_{\mathcal Q}^2$, and matching onto pure DGLAP evolution for scales below $m_{\mathcal Q}^2$. However, in the $Q^2$ range presently accessible in experiment, these logarithms appear to be under control, as indicated by the good agreement with the data up to the largest virtualities, see Fig.~\ref{fig:chris_1}. Their resummation might only become necessary at higher scales, and in any case beyond the kinematic reach of the EIC. Current data are nonetheless statistically limited in the moderate-to-large $Q^2$ region; the EIC will therefore provide essentially the first comprehensive coverage over a wide range, yielding sufficient statistics to determine whether resummation may already be relevant after all. Other numerical effects in CF such as the so-called `$Q_0$ subtraction'~\cite{Jones:2016ldq}, crucial for a fruitful description of the $J/\psi$ photoproduction data~\cite{Flett:2019pux, Flett:2020duk, Flett:2021fvo, Flett:2022ues}, are expected to be numerically less important for $J/\psi$ electroproduction in the kinematic region $Q^2 \gg m_c^2$. \textcolor{black}{This is because} the corresponding power correction $\mathcal O(Q_0^2/\mu_F^2)$, where $Q_0$ is the PDF input scale and $\mu_F^2 = \mathcal O(Q^2 + 4m_c^2)$, is no longer of $\mathcal O(1)$. 

\section{Conclusions}
\label{conc}
In this work, we have applied the NLO coefficient functions for the exclusive electroproduction of heavy vector mesons, computed in our earlier work~\cite{Flett:2021ghh}, to perform detailed phenomenological studies of electroproduction in electron-proton collisions, relevant for measurements at HERA and the upcoming EIC. Our results, when compared against the existing HERA measurements, demonstrate that the NLO CF framework provides a sound description of the available cross sections over a wide range of photon virtualities $Q^2$ and centre-of-mass energies $W$. This lends confidence to the robustness of the formalism and its relevance for current and future phenomenological applications. The forthcoming EIC will open a new era of precision studies in this sector, allowing for much more stringent tests of the theoretical description. In anticipation of these measurements, we have provided quantitative predictions that will serve as valuable benchmarks for future analyses.

\sloppy
Our results highlight the importance of logarithmically-enhanced terms of order $(\alpha_s \ln^2(Q^2/m_{\mathcal Q}^2))^n$ in the large-$Q^2$ regime. To our knowledge, a dedicated and systematic resummation framework tailored to heavy vector meson electroproduction at large $Q^2$ for such a tower of logarithms is not yet fully established. A detailed implementation of the matching procedure between the IREE and DGLAP formalisms discussed in the text to address this is left for future work. Such improvements could provide for more favourable data comparisons of exclusive heavy vector meson electroproduction at the EIC, as well as at the future FCC-eh~\cite{FCC:2018byv} or LHeC~\cite{LHeC:2020van} facilities, where the reach in $Q^2$ will be greater still.

Such studies all play a role in deepening our understanding of nucleon structure and the dynamics of strong interactions in exclusive processes, enabling increasingly precise access to the underlying partonic structure of the nucleon across multiple scales. 

\section*{Acknowledgements}
I would like to thank M.~G. Ryskin for useful discussions. I am supported by the Marie Skłodowska-Curie Action (``AutomOnium''), funded by the European Union under grant agreement No. 101204057. I also acknowledge support from the Agence Nationale de la Recherche (ANR) via the grant ANR-20-CE31-0015 (``PrecisOnium''), as well as
the IDEX Paris-Saclay ``Investissements d'Avenir'' (ANR-11-IDEX-0003-01) through the GLUODYNAMICS project funded by the ``P2IO LabEx (ANR-10-LABX-0038)'', the French CNRS via the IN2P3 projects ``GLUE@NLO” and ``QCDFactorisation@NLO'', and the COPIN-IN2P3 project \#12-147 ``kT factorisation and quarkonium production in the LHC era''.

\bibliographystyle{unsrt}
\bibliography{references.bib} 
\end{document}